# Revisiting ϵ Eridani with NEID: Identifying New Activity-sensitive Lines in a Young K Dwarf Star

Sarah Jiang[1], Arpita Roy[1,2], Samuel Halverson[3], Chad F. Bender[4], Carlos Selgas[5], O. Justin Otor[1], Suvrath Mahadevan[6,7], Guðmundur Stefánsson[8,11], Ryan C. Terrien[9], and Christian Schwab[10]
[1] Space Telescope Science Institute, 3700 San Martin Drive, Baltimore, MD 21218, USA
[2] Department of Physics and Astronomy, Johns Hopkins University, 3400 N. Charles Street, Baltimore, MD 21218, USA
[3] Jet Propulsion Laboratory, 4800 Oak Grove Drive, Pasadena, CA 91109, USA
[4] Steward Observatory, The University of Arizona, 933 N. Cherry Avenue, Tucson, AZ 85721, USA
[5] Eberly College of Science, 517 Thomas Building, The Pennsylvania State University, University Park, PA 16802, USA
[6] Department of Astronomy & Astrophysics, 525 Davey Laboratory, The Pennsylvania State University, University Park, PA 16802, USA
[7] Center for Exoplanets and Habitable Worlds, 525 Davey Laboratory, The Pennsylvania State University, University Park, PA 16802, USA
[8] Department of Astrophysical Sciences, Princeton University, 4 Ivy Lane, Princeton, NJ 08540, USA
[9] Carleton College, One N. College Street, Northfield, MN 55057, USA
[10] Department of Physics and Astronomy, Macquarie University, Balaclava Road, North Ryde, NSW 2109, Australia
*Received 2022 November 21; revised 2023 October 27; accepted 2023 October 27; published 2023 December 6*

## Abstract

Recent improvements in the sensitivity and precision of the radial velocity (RV) method for exoplanets have brought it close, but not quite to, the threshold ($\sim 10$ cm s$^{-1}$) required to detect Earth-mass and other potentially habitable planets around Sun-like stars. Stellar activity-driven noise in RV measurements remains a significant hurdle to achieving this goal. While various efforts have been made to disentangle this noise from real planetary signals, a greater understanding of the relationship between spectra and stellar activity is crucial to informing stellar activity mitigation. We use a partially automated method to analyze spectral lines in a set of observations of the young, active star ϵ Eridani from the high-precision spectrograph NEID, correlate their features (depth, FWHM, and integrated flux) with known activity indicators, and filter and curate for well-defined lines whose shape changes are sensitive to certain types of stellar activity. We then present a list of nine lines correlated with the S-index in all three line features, including four newly identified activity-sensitive lines, as well as additional lines correlated with the S-index in at least one feature, and discuss the possible implications of the behavior observed in these lines. Our line lists represent a step forward in the empirical understanding of the complex relationships between stellar activity and spectra and illustrate the importance of studying the time evolution of line morphologies with stabilized spectrographs in the overall effort to mitigate activity in the search for small, potentially Earth-like exoplanets.

*Unified Astronomy Thesaurus concepts:* Stellar activity (1580); Spectral line lists (2082); Radial velocity (1332); Stellar spectral lines (1630); Exoplanet astronomy (486)

*Supporting material:* machine-readable tables

## 1. Introduction

The radial velocity (RV) method is a vital tool for detecting exoplanets by measuring and analyzing the Doppler shifts in stellar spectra caused by a planet "wobbling" its host star. The RV method has seen vast improvements in precision in recent years, with current-generation instruments (e.g., ESPRESSO, EXPRES, MAROON-X, NEID, KPF) aiming for precisions of less than 1 m s$^{-1}$ (Schwab et al. 2016; Gibson et al. 2018; Blackman et al. 2020; Seifahrt et al. 2020; Pepe et al. 2021). However, as instruments near the 10 cm s$^{-1}$ precision threshold for detection of Earth-like planets around Sun-like stars, the detection of these planets is still limited primarily by the problem of stellar activity signals contaminating RV measurements and mimicking planetary signals (National Academies of Sciences, Engineering, & Medicine 2018; Robertson et al. 2014; Sairam & Triaud 2022). Developing methods for identifying and removing stellar activity from these measurements is a crucial step in achieving the much sought-after goal of detecting Earth analogs with extreme precision RV (EPRV).

Recent work has made significant advances in applying various techniques to characterizing and mitigating the effect of stellar activity on RV measurements. These techniques include employing regular-cadence observing strategies (Jeffers et al. 2022), using proxies and models to estimate the effect of stellar activity and subtract it from RV measurements (Ervin et al. 2022; Haywood et al. 2022), and applying various statistical techniques to the spectra or cross-correlation functions (CCFs) to identify and separate variation caused by different sources of activity (Cretignier et al. 2022; Siegel et al. 2022; Simola et al. 2022; Zhao et al. 2022). In addition to these techniques, machine learning has the potential to become a promising new avenue for astronomical data analysis. In the realm of EPRV, de Beurs et al. (2022) used a convolutional neural network (CNN) to correlate shape changes in the CCFs of both simulated stellar data and HARPS-N solar data with stellar activity and removed these stellar activity signals from the measured RVs. However, the use of machine learning in the field of EPRV is still in its infancy. While recent work has shown the potential of machine learning to be a powerful tool in identifying and parsing stellar activity signals from real planetary signals, more work is needed to understand the physical relationships between stellar activity and changes in the features of stellar spectra in order to effectively deploy both conventional and machine learning techniques on these spectra.

---

[11] Henry Norris Russell Fellow.

Wise et al. (2018; henceforth referred to as "W18") used a method of correlating line depths/core fluxes with well-known activity indicators such as Ca II H & K/S-index (Isaacson & Fischer 2010) to identify a list of activity-sensitive lines in archival HARPS spectra of $\epsilon$ Eridani ($\epsilon$ Eri) and $\alpha$ Centauri B ($\alpha$ Cen B). These 42 lines spanning HARPS's wavelength range of 3780–6910 Å (Pepe et al. 2000), with features that exhibit periodicity at the star's rotation period, are a useful tool for investigating how certain lines are tied to stellar activity. For example, Lisogorskyi et al. (2020) used the line list published by W18 as a reference to help identify active regions of the spectra in order to explore how the removal of these regions helps improve RV calculations. The identification of activity-sensitive lines is therefore a first step toward uncovering the complex relationships between spectral features and stellar activity, and eventually disentangling this information from the RVs.

We expand upon the work of W18 to develop a partially automated method using Python to explore periodic line shape behaviors in the NEID spectra. NEID is a high-precision Doppler spectrograph installed on the 3.5 m WIYN Telescope[12] at Kitt Peak National Observatory, with a baseline RV precision of $\sim$30 cm s$^{-1}$ (Schwab et al. 2016). In addition to taking advantage of NEID's broader wavelength range (3800–9300 Å) to expand the list of activity-sensitive lines into redder wavelengths, we also investigate several other line features along with the line depth (specifically, FWHM and integrated flux) and correlate with multiple activity indices and RV to further explore and verify the sensitivity of our lines to multiple parameters.

We use NEID observations of $\epsilon$ Eri spanning a 6 month period from 2021 September to 2022 February, reduced with the NEID Data Reduction Pipeline (DRP) Version 1.1.2[13], to identify the best candidates for our line lists. We chose $\epsilon$ Eri as our study target since it is a young and active star in which we expected to observe significant levels of stellar activity, which would also allow us to verify the lines published by W18 using HARPS data of $\epsilon$ Eri. While we only investigate NEID data of $\epsilon$ Eri in this paper, we designed our method and code for extracting and correlating line features with activity indices and RV with flexibility in mind, so that this method can be adapted and applied to other data from different instruments and/or stars.

We present a list of nine lines, including four newly identified lines, that show correlations in several different line features with the S-index, as well as additional line lists that correlate with the S-index in single or double features, each of which offers a new and extended window into measuring and analyzing stellar activity using activity-sensitive lines in the spectra. We produced these lists using a partially automated method combining first our code for correlating line parameters and then a manual analysis to remove the lines for which the data displayed inaccurate fits or outliers. The results are curated lists of well-defined, vetted lines that can be used as reference points for future work probing information about stellar activity in the spectra and understanding the physics of stellar activity-sensitive changes in spectral features.

In this paper, we begin by detailing our data processing and the development of our Python code for performing our line analysis and correlation in Section 2. Then, in Section 3, we describe first our verification of the line list published by W18, then describe how we determined our new line lists and the criteria we used to further curate our line lists. Finally, in Section 4, we present our conclusions and discuss how our results can help inform further research in stellar activity and improve the precision of the RV method for detecting exoplanets.

## 2. Methods

### 2.1. Data processing

Our analysis begins with a set of observations of a star (in this case, $\epsilon$ Eri) and any line list or set of line lists. We use Level 2 data files reduced by the NEID DRP Version 1.1.2, which are multiextension `astropy` Flexible Image Transport System (FITS) files that contain the spectroscopic data (`SCIFLUX`) for each observation as well as various metadata, calibration information, blaze functions for each spectral order, measured RVs, and activity indicators. The median extracted signal-to-noise ratio (S/N) for our $\epsilon$ Eri observations is 463, so for this analysis, we first filter our observations for S/N of >300 to exclude three low-S/N observations. While processing our data, we also noticed a trend in the NEID $\epsilon$ Eri data in which certain observations with high photon counts seemed to near the saturation threshold of NEID and became "flattened," similar to what one would expect in a nonlinear response of the CCD, thus depressing the spectra and consequently affecting the shape of the spectral lines in those regions. We manually removed the observations affected by this phenomenon for this analysis. Then, before performing line-by-line analysis, we Doppler shift each spectrum to the rest frame of the star using the NEID DRP-derived radial velocity (`CCFRVMOD`) and barycentric velocity (`SSBRVORD` where `ORD` is the corresponding NEID echelle order number) to ensure consistency in our automated analysis. We convert this combined velocity to a wavelength shift and subtract this from the wavelength axis (`SCIWAVE`) of each spectrum. Then, we divide the spectra by the blaze function of each order (`SCIBLAZE`). After correcting for the blaze function, we normalize the spectra by fitting a segmented polynomial over sections of the spectra to estimate the continuum and dividing the spectra by this estimate. The resulting blaze-corrected, continuum-normalized, and shifted spectra are the inputs to our partially automated line-by-line analysis. Throughout this paper, all wavelengths reported are wavelengths in a vacuum and in the stellar rest frame.

To identify our initial candidates for activity-sensitive lines, we investigated 37 NEID observations for $\epsilon$ Eri spanning a range of 6 months from 2021 September to 2022 February, since it is a young and active star whose spectra had shown previous sensitivity to activity in W18. After applying the data processing detailed above, we were left with 32 high-S/N observations of $\epsilon$ Eri for our analysis. We used four different line lists to perform our analysis: a list of known, nonmagnetically sensitive solar lines[14] from Altrock et al. (1975;

---

[12] The WIYN Observatory is a joint facility of the University of Wisconsin–Madison, Indiana University, NSF's NOIRLab, the Pennsylvania State University, Purdue University, and the University of California, Irvine.
[13] https://neid.ipac.caltech.edu/docs/NEID-DRP/
[14] Defined by Altrock et al. (1975) as having Landé factor = 0.

Table 1
Noise Floors

| Parameter | Altrock RV $\|\rho\|$ | Altrock S-index $\|\rho\|$ | Altrock RV $\|\tau\|$ | Altrock S-index $\|\tau\|$ | Parameter | VALD RV $\|\rho\|$ | VALD S-index $\|\rho\|$ | VALD RV $\|\tau\|$ | VALD S-index $\|\tau\|$ |
|---|---|---|---|---|---|---|---|---|---|
| Centroid | 0.12 | 0.08 | 0.13 | 0.14 | Centroid | 0.12 | 0.19 | 0.11 | 0.13 |
| Depth | 0.19 | 0.20 | 0.20 | 0.30 | Depth | 0.19 | 0.34 | 0.15 | 0.17 |
| FWHM | 0.11 | 0.14 | 0.08 | 0.11 | FWHM | 0.26 | 0.34 | 0.14 | 0.18 |
| Integrated Flux | 0.33 | 0.37 | 0.26 | 0.34 | Integrated Flux | 0.17 | 0.25 | 0.13 | 0.14 |

**Note.** The uncertainties on these noise floor correlation coefficients range from 4 to 6 orders of magnitude smaller than the coefficients.

henceforth referred to as "Altrock et al."), the ESPRESSO K2 mask, a line list created by compiling several ESPRESSO masks for different star types (M2, K2, K6, G2, G8, G9, and F9) and removing duplicates (lines that are within 0.2 Å of each other, a window that is defined in Section 2.3) between masks, and an empirical line list created by our own line-finding algorithm applied to NEID solar spectra. We also used a line list generated by the Vienna Atomic Line Database 3 (VALD3, henceforth referred to as "VALD"; Pakhomov et al. 2019) using the stellar parameters of $\epsilon$ Eri from Drake & Smith (1993) and a line detection threshold of 0.1 to explore various features and atomic properties of our lines. We used this VALD line list and the nonmagnetically sensitive lines to calculate an expected noise floor for uncorrelated lines (see Section 2.2). The other three lists were combined into one line list in an effort to build as complete as possible of a list of all lines in the spectra: the ESPRESSO K2 mask served as a reference for $\epsilon$ Eri, the combined ESPRESSO mask list allowed us to probe for any lines that the K2 mask alone might have missed in the NEID $\epsilon$ Eri spectra, and finally, the empirical line list allowed us to probe the wavelength region beyond 7880 Å (the reddest wavelength detected by ESPRESSO) that is available to us with NEID spectra. We also separately investigated the list of activity-sensitive lines published by W18 (henceforth referred to as the "Wise line list") to verify their results.

### 2.2. Calculating Noise Floor

Noisy, highly variable lines should evince significantly higher values than the noise floor. We investigate two metrics for the definition of "quiet" lines to calculate an expected noise floor for uncorrelated lines: (1) known, nonmagnetically sensitive solar lines from Altrock et al., and (2) low excitation energy lines from VALD.

We first calculated correlation matrices for each line, then computed an error-weighted mean of each individual correlation across all matrices to generate a "master" noise floor correlation matrix. Thus, these correlation coefficients are the values that we expect, on average, non-activity-sensitive lines to exhibit for each parameter activity index combination, purely from noise, and act as reference thresholds for values we expect to indicate significant correlation. We computed these noise floors using the absolute values of both Pearson's correlation coefficient ($\rho$) and Kendall's $\tau$ coefficient ($\tau$). While we calculated the noise floor for all measured activity indices, we ultimately chose to investigate only the relationship between each line parameter and S-index (see Section 3.2). We therefore display the noise floors for correlation with RV and S-index only, including the errors on each correlation coefficient, in Table 1.

To cross-check metrics of quietness, we compared the lowest variability lines in our empirical sample (see Section 2.3), the magnetically quiet lines from Altrock et al., and the lowest excitation energy lines in VALD. We find that (a) only four out of 13 of Altrock et al.'s quiet lines had VALD excitation energies that were lower than the mean, (b) only one of our 50 empirically lowest variability lines overlapped with Altrock et al.'s list, (c) Altrock et al.'s sample has no overlap with the 50 lowest excitation energies from VALD, and (d) our 50 empirically lowest variability lines has no overlap with the 50 lowest excitation energies from VALD.

The results of these analyses suggest that Altrock et al.'s quiet lines are not the same lines that we found to have the lowest variability in observed line parameters and/or excitation energies from VALD. While noise floors computed from the Altrock et al. sample and VALD are different, we include both thresholds in Table 1 for completeness.

### 2.3. Line Analysis

For each line in the compiled line list, our code retrieves all observations, performs the normalization and shifting mentioned in Section 2.1, extracts the RV and all measured activity indices[15] (Ca II H & K/S-index, He I D3 line, Na I D1/D lines, H$\alpha$, Ca I, Ca II infrared triplet, Na I near-infrared doublet, and Paschen delta) for each observation, then finds the line in the spectrum and fits and measures the centroid, depth, FWHM, and integrated flux across the line. We fit each line with a Gaussian function using `scipy.curve_fit` within an initial window of $\pm 0.2$ Å around the line center. In order to ensure that the Gaussian function is accurately fitting the line, if the centroid of the initial fit is not within 0.01 Å of the line center, we gradually narrow the window until it is. We discard lines for which the centroid of the fit never reaches within 0.01 Å of the line center at any window within $\pm 0.2$ Å, as we are searching for well-defined lines that can be identified and used as references in other observations. We measure the integrated flux by defining a uniform $\pm 3$ km s$^{-1}$ window around each line center, interpolating the flux over that window, and summing the interpolation. Figure 1 shows an example line and how each parameter is measured. Note that the window over which we calculate integrated flux does not encompass the full line. Our tests showed that using a window of $\pm 3$ km s$^{-1}$ to calculate integrated flux captures enough variability in the flux to be sensitive to shape changes without sacrificing sensitivity by including parts of the wings of each line; i.e., wider windows were less sensitive to our activity indicators. The exception to this were lines that exhibited sensitivity to activity in FWHM and integrated flux only, but not depth (see Section 3). This is most likely due to the fact that integrated flux variability in those lines is driven *by* FWHM variation, in which case a wider window is more sensitive to that variation. However, in this case, using the window of $\pm 3$ km s$^{-1}$ acts as a filter in which only lines that exhibit FWHM variation strong enough to affect the narrower window are detected as candidates for activity-correlated lines.

---

[15] For information on how the NEID DRP Version 1.1.2 calculates these activity indices, see here: https://neid.ipac.caltech.edu/docs/NEIDDRP/algorithms.html#stellar-activity-info.

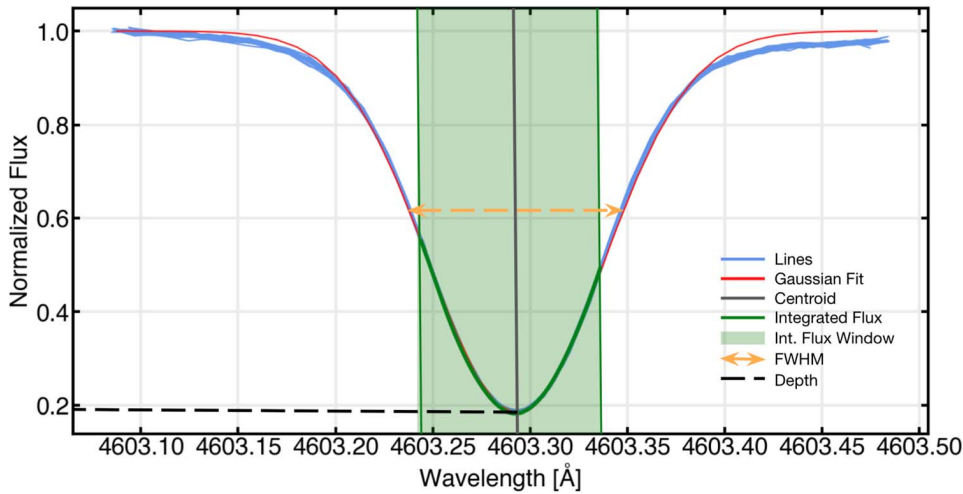

**Figure 1.** Example of how we fit each line (blue) with a standard Gaussian function (red) and extract all line parameters. The centroid, depth, and FWHM of the line are measured from the Gaussian fit. The range over which we calculate integrated flux in this example line is denoted in green. Note that the exact wavelength bounds of this region vary slightly from line to line since our integrated flux window is uniform in velocity, not wavelength, though on average the integrated flux window captures a similarly wide region of each line around the line core as shown in this figure.

After fitting for and extracting all line features, we save these features along with the measured RV, activity indices, and errors on all parameters into an xarray DataArray object ("cube"). The resulting cube holds the following parameters and corresponding errors: centroid, depth, FWHM, integrated flux, RV, and all measured activity indices for each line and observation. This cube is available on GitHub[16] with a copy deposited in Zenodo.[17]

The data can then be explored by retrieving any desired line feature or activity index for any line and observation by indexing the cube. The next step in our analysis involves iterating through the cube and correlating each feature with each activity index. However, some of the measured activity indices extracted from the observations were calculated multiple times using different procedures. Before performing the correlations, we combine each of these repeated activity indices into one using an error-weighted mean.

For each line in the cube, we compute a correlation matrix of each line feature (centroid, depth, FWHM, and integrated flux) with the measured RV and combined activity indices (including RV, nine indices in total) for a total of 36 separate correlation coefficients. We calculate these matrices using the absolute values of both $\rho$ and $\tau$. We compute errors on each of these correlation coefficients by calculating the same correlation coefficient within the error bounds of each parameter being correlated (e.g., depth and S-index) and taking the standard deviation of the resulting distribution. We then determine thresholds for correlation in both $\rho$ and $\tau$ (this process is detailed in Section 3) and use these to filter through these correlation matrices for lines that we find to be correlated in various parameters.

### 2.4. Removing Outliers and Telluric-contaminated Lines

After our initial filter for activity-sensitive lines, we manually cleaned each list for outliers and telluric-contaminated lines. First, we pruned lines from each line list that either exhibited clear outliers in the plots of their features correlated with the S-index (such that the slope of the correlation was significantly skewed) or where the lines themselves visually exhibited two or more peaks within in the fitting window or were extremely noisy (such that the Gaussian fit did not accurately fit and measure the line's parameters). Examples of the visual criteria we used to prune our line lists are shown in Figure 2. This pruning serves as a vetting of the lines we analyze for only lines that are well defined over time.

Telluric absorption features caused by Earth's atmosphere tend to vary both in position and strength relative to stellar features. To ensure that our selection of activity-sensitive lines is not affected by these variations, we compared the integrated fluxes of each line to the integrated flux computed across the same window on the telluric model included in the NEID FITS files (TELLURIC). In the current NEID DRP, the telluric model is not normalized to 1, so we first renormalize the telluric model. We then computed an additional "baseline" model: a horizontal line at the maximum value of the telluric model (1 after normalization). Then, for each line, we calculated the absolute value of the difference between the baseline and the actual integrated flux of the line, as well as the absolute value of the difference between the baseline and the integrated flux over the telluric model in the same window. These values represent the deviation from the baseline of the actual observed spectrum and the deviation from the baseline of the telluric model. By dividing the telluric deviation by the observed spectrum deviation, we obtained a "percentage of telluric contamination" for each line. We then used a threshold of 1% to filter for telluric-contaminated lines and removed these lines from our line lists.

---
[16] https://github.com/sarahxj/eps_eri_cube
[17] doi: 10.5281/zenodo.10085919.

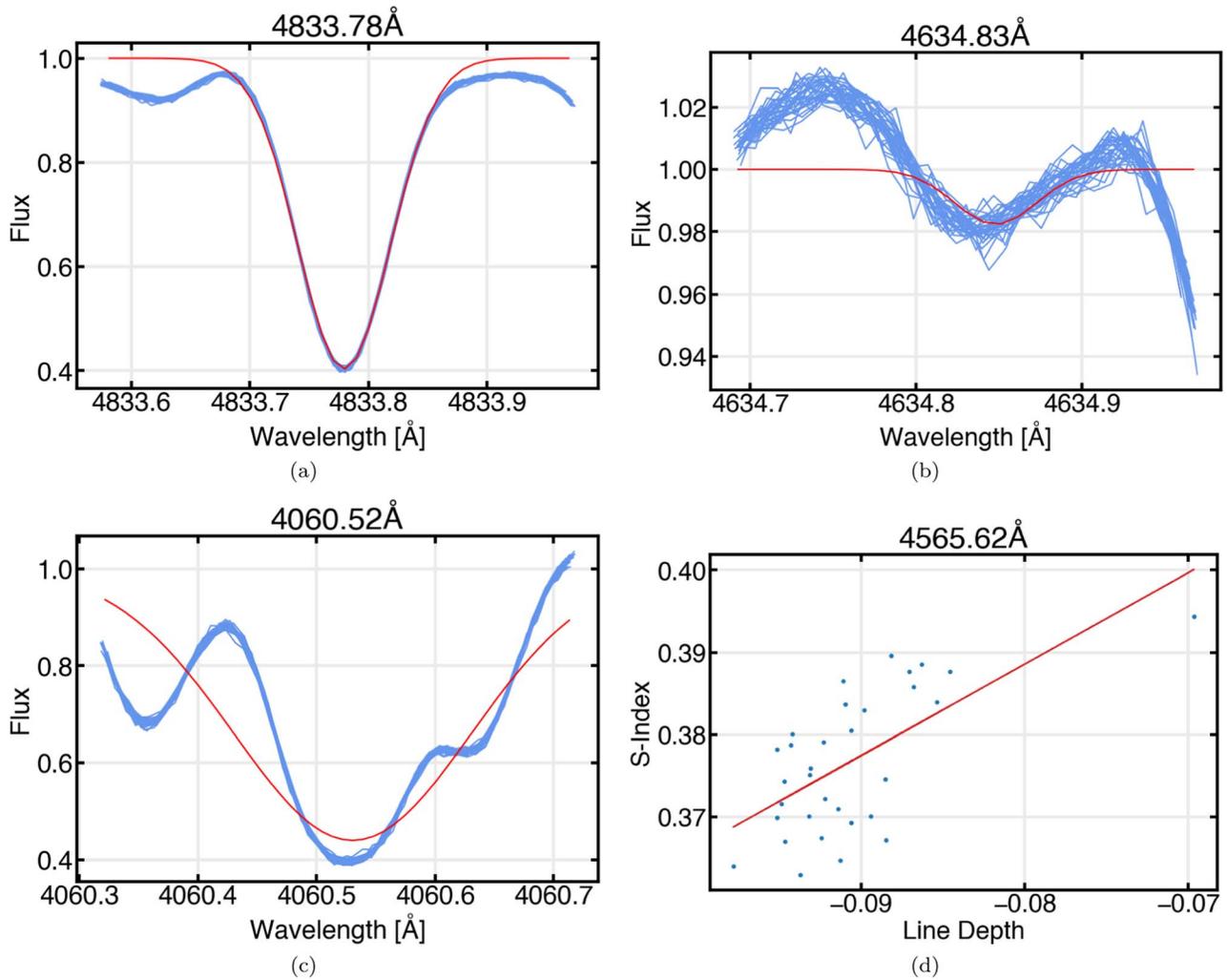

**Figure 2.** Examples of the visual criteria we used to prune our line lists. We determine whether or not to remove a line based on whether the Gaussian fit appears to accurately fit and measure a line's parameters, and if the correlations appear to exhibit a significant outlier that skews the correlation. (a) An example of a "good" line that we do not remove, since the Gaussian fit appears to match the line within its wings. (b) An example of a noisy line such that the Gaussian fit does not accurately fit the line. (c) An example of a blended line with multiple peaks in the fitting window, such that the Gaussian fit does not accurately fit the line. (d) An example of a line whose plot correlating line depth with the S-index exhibits a significant outlier that skews the best-fit line.

## 3. Results

### 3.1. Verification of Wise Line List

The first step in our analysis was to investigate and verify the Wise line list to both check our methods against previously studied lines and verify the lines published by W18. These 42 lines were found by W18 to have line core fluxes (the flux at the absolute minimum of the line) that correlated with the S-index ($|\tau| \geqslant 0.5$) and were periodic at $\alpha$ Cen B's rotation period, suggesting that these lines trace activity similarly to known activity indices. W18 also investigated the "half-depth range" and "center of mass," which are defined similarly to FWHM and line centroid, and also reported lines whose correlations between half-depth range and S-index were $\geqslant 0.5$, even if their correlations between line core flux and S-index were not. These additional lines still seemed to probe photospheric effects on the spectra, if not as strongly. While these lines did not meet the 0.5 threshold of correlation between line core flux and S-index, all of them did exhibit correlations of at least 0.4. Accordingly, we adopted a threshold of $|\tau| = 0.4$ to determine whether a line was correlated or not in our parameters. We employ $\tau$ as one of our measures of correlation for similar reasons as W18: it probes both the covariance of our parameters with the S-index and the intrinsic measurement noise and additionally allows a direct comparison between our measurements and the results from W18. For robustness, we also employ Pearson's correlation coefficient $\rho$, a commonly used standard for measuring linear correlation between two variables, as a second metric. Pearson's correlation coefficient is more sensitive to outliers but incorporates the magnitudes of the variables themselves, which makes it a strong indicator of linear correlation between continuous variables, while Kendall's $\tau$ coefficient is a rank coefficient but is less susceptible to outliers as a result. We used a correlation threshold of $|\rho|=0.5$ since tests showed that most lines that exhibited $|\tau| \geqslant 0.4$ also exhibited $|\rho| \geqslant 0.5$, and vice versa, but some did not. Therefore, filtering for lines that meet both thresholds adds an additional layer of confidence in the lines' correlations.

These thresholds were chosen based on these previous results and used to verify the activity-sensitive lines published by W18, but they are also $>3\sigma$, where $\sigma$ is the error on the noise floor correlations, above the respective $|\rho|$ and $|\tau|$ noise floor values for correlations between each of our line features (depth, FWHM, and integrated flux) and S-index; see Table 1. We therefore chose to employ these same thresholds for our independent line analysis. While we found that some of the noise floor values for correlations with other activity indices (e.g., for correlation between FWHM and Hα) were $\geqslant 0.4$ or $\geqslant 0.5$, respectively, we eventually chose to focus only on correlations between line parameters and S-index, a decision that is detailed in Section 3.2. With that in mind, we expected our analysis of NEID ε Eri spectra to show $|\tau| \geqslant 0.4$ and $|\rho| \geqslant 0.5$ for correlations between depth (analogous to line core flux) and S-index for all lines in the Wise line list.

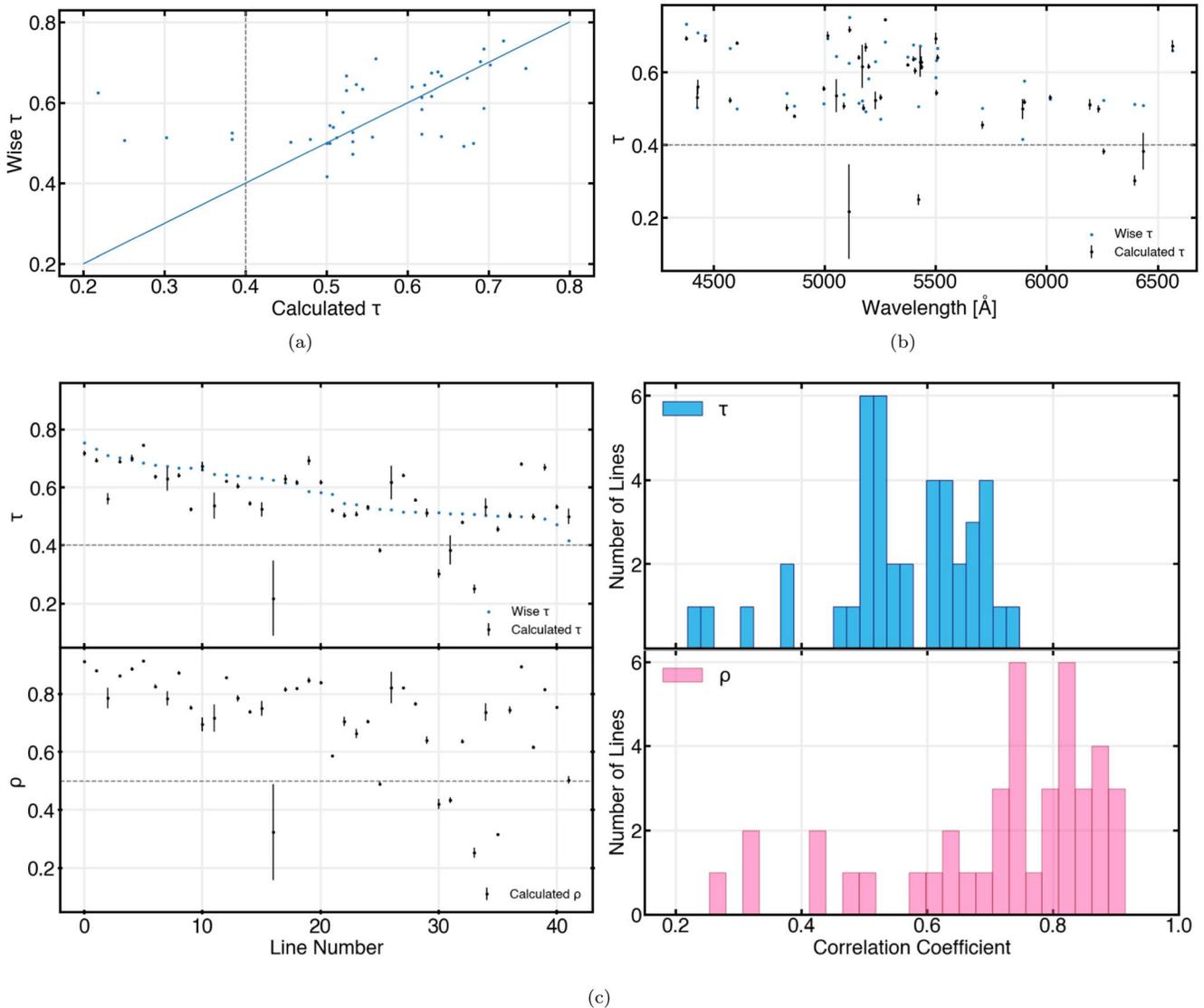

**Figure 3.** Results of our analysis of the Wise line list using our partially automated method and NEID ε Eri spectra. In these graphs, the $|\tau|$ and $|\rho|$ values are for the correlation between line core flux/depth and S-index. (a) $|\tau|$ values for each line in the Wise line list calculated by Wise et al. (2018; "W18") plotted against the $|\tau|$ values we calculated for the same lines using NEID data of ε Eri. The 1:1 line is marked in blue, and our threshold for correlation in $|\tau|$ is marked in gray. Five lines did *not* meet our threshold for correlation in $|\tau|$. (b) Both our and W18's $|\tau|$ values for each line in the Wise line list, organized by wavelength. Our threshold for correlation in $|\tau|$ is marked in gray. (c) Both our and W18's $|\tau|$ values and our $|\rho|$ values (W18 did not calculate $|\rho|$) for each line in the Wise line list, organized by line number in descending order of $|\tau|$ values calculated by W18, and the corresponding histograms. Our thresholds for correlation in $|\tau|$ and $|\rho|$ respectively are marked in gray. In addition to the lines that did not meet our threshold for correlation in $|\tau|$, one line was found to meet our threshold for correlation in $|\tau|$, but not $|\rho|$. The corresponding histograms show the $|\tau|$ and $|\rho|$ values we calculated for correlations between line depth and S-index for the Wise line list. $|\rho|$ values are consistently higher than corresponding $|\tau|$ values for the same line.

The results of our analysis of the Wise line list using our partially automated analysis and NEID ε Eri spectra are shown in Figure 3. For each individual line, we did not expect our calculated values of $|\tau|$ for correlation between depth and S-index to necessarily match the values found by W18, since we are looking at a different time period of ε Eri data in which activity levels were slightly lower: W18 observed an S-index range of 0.388–0.469 in their archival HARPS data, while our data spanned an S-index range of 0.363–0.394. This small difference might arise from the details of our measurement method or due to true subtle variations from long-term magnetic cycles on the star. However, we did see that for most lines in the Wise line list, our calculated $|\tau|$ values for correlation between depth and S-index were $\geqslant 0.4$ except for five lines: 5108.87 Å (Fe I), 6254.29 Å (Fe I), 6395.38 Å (Fe I), 6432.63 Å (Fe I), and 5421.86 Å (Mn I). These five lines also did not exhibit $|\rho|$ values $\geqslant 0.5$, as well as one additional line, 5708.58 Å (V I), that had $|\tau| > 0.4$ but $|\rho| < 0.5$. Overall, we found that 37 of the 42 lines published by W18 met the criterion of having a value of $|\tau|$ for correlation between depth and S-index $\geq 0.4$ (both our threshold for correlation with $|\tau|$ and the threshold used in W18), while 36 of the 42 lines met both of our criteria ($|\tau| \geq 0.4$ and $|\rho| \geq 0.5$) for correlation between depth and S-index.

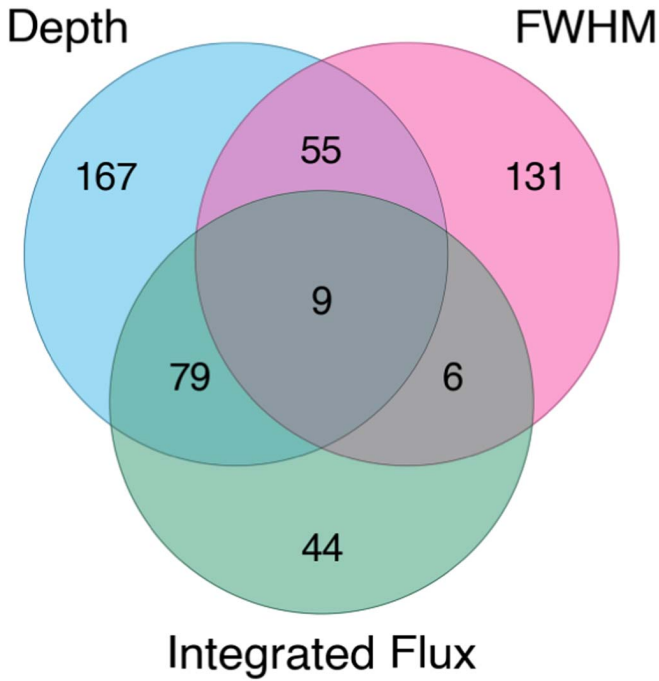

**Figure 4.** Venn diagram of our seven line lists and the number of lines in each list. All numbers/line lists are exclusive; i.e., we found nine lines correlated with the S-index in all three parameters (depth, FWHM, and integrated flux), 55 lines correlated with the S-index in depth and FWHM only, 79 lines correlated with the S-index in depth and integrated flux only, six lines correlated with the S-index in FWHM and integrated flux only, and 167, 131, 44 lines respectively correlated with the S-index in depth, FWHM, or integrated flux only.

In addition, we include histograms of the $|\tau|$ and $|\rho|$ values we calculated for the correlations between depth and S-index for the Wise line list in Figure 3. The $|\rho|$ values are consistently higher than the $|\tau|$ values for the same correlation in the same line, which supports our choice to use a lower correlation threshold for $|\tau|$ ($\geq 0.4$) than $|\rho|$ ($\geq 0.5$).

### 3.2. New Line Lists

To independently identify our new list(s) of activity-sensitive lines, we first combined all three lists described in Section 2 (the ESPRESSO K2 mask, the combined ESPRESSO mask list, and the empirical line list created from NEID spectra) and removed duplicates to produce a single "master" list of lines containing ∼11,500 lines total. We then performed our line-by-line analysis on this master list and using our set of 32 NEID observations of $\epsilon$ Eri. This resulted in a data cube holding all parameters and measured activity indices for each of our ∼11,500 lines and 32 observations. Using this cube, we then computed the corresponding correlation matrices for each line.

We then filter these correlation matrices for the thresholds that we determined to indicate significant correlation ($|\tau| \geq 0.4$ and $|\rho| \geq 0.5$). We expect stellar activity to be more likely to cause shape changes in the lines rather than translational shifts. While stellar activity can sometimes cause shape changes that masquerade as translational shifts in the line centroid depending on the method used to measure the centroid, correlating only shape changes with stellar activity avoids the possibility of misinterpreting fits skewed by shape changes for real translational shifts. Moreover, we know that real planetary signals should cause translational shifts but not shape changes. Therefore, we focus on the correlations between depth, FWHM, and integrated flux with activity indices to identify our activity-sensitive lines. We began by filtering only for correlations between depth, FWHM, and integrated flux with the S-index, since the S-index is a widely known and used activity indicator. We filtered the correlation matrices for lines that had $|\tau| \geq 0.4$ and $|\rho| \geq 0.5$ for correlations between any of these three parameters with the S-index. This yielded three lists of correlated lines: one that exhibited $|\tau| \geq 0.4$ and $|\rho| \geq 0.5$ for correlation between depth and S-index, one that exhibited $|\tau| \geq 0.4$ and $|\rho| \geq 0.5$ for correlation between FWHM and S-index, and one that exhibited $|\tau| \geq 0.4$ and $|\rho| \geq 0.5$ for correlation between integrated flux and S-index.

We also tried filtering for correlations between each line feature and our other activity indices. While we did find that some lines showed significant correlations between line features and other activity indices (such as H$\alpha$), we found that these lines generally did not overlap with the lines whose line features were correlated with the S-index. This could be due to several factors: the lines that we filtered for (well-defined lines that could be fit within a window of $\pm 0.2$ Å) could be affected mostly by the particular chromospheric and photospheric changes reflected in the S-index in $\epsilon$ Eri. Alternatively, the line features that we selected (depth, FWHM, and integrated flux) themselves could be more sensitive to S-index than other activity indices. Since different activity indicators trace different sources of stellar activity, we chose to focus on the lines that we found to be correlated with the S-index for the scope of this paper, as S-index is the most widely known and used activity indicator in stellar activity studies (Wilson 1968; Vaughan et al. 1978) and serves as a useful reference point for probing spectral sensitivity to stellar activity. We encourage future studies to explore lines sensitive to stellar activity through other indices.

After compiling our three lists of lines correlated with the S-index separately in depth, FWHM, or integrated flux, we then cross referenced each of these three lists with the others, as well as all three with each other, to produce four additional lists: lines that exhibited correlation with the S-index in both depth and FWHM, depth and integrated flux, and FWHM and integrated flux, and lines that exhibited correlation with the S-index in all three parameters. Then we removed duplicates from each list, such that our final seven lists are exclusive and do not overlap with each other. Finally, we manually pruned outliers and telluric-contaminated lines from the lists, as detailed in Section 2.4.

This resulted in seven exclusive line lists: lines correlated with the S-index in all three parameters (depth, FWHM, and integrated flux), lines correlated with the S-index in depth and FWHM only, lines correlated with the S-index in depth and integrated flux only, lines correlated with the S-index in FWHM and integrated flux only, and lines correlated with the S-index in depth, FWHM, or integrated flux only. These lists had 9, 55, 79, 6, 167, 131, and 44 lines respectively. Although the number of activity-sensitive lines we identified (even just in depth only) is significantly higher than the 42 lines identified by W18, we note that we investigated a list of ~11,500 lines while W18 looked at ~1500 lines, a difference of nearly an order of magnitude. Figure 4 displays the number of lines in each list and how they overlap. Figure 5 shows the range of each line list over NEID's wavelength range; note that the line lists are fairly evenly distributed below ~6500 Å. While we found a couple of correlated lines in the redder half of NEID's wavelength range, there is a gap between ~6500 and 9000 Å. This may be due to physical effects that make lines in that range less sensitive to chromospheric variation in the features that we chose, the presence of large telluric water and oxygen absorption bands in this region, biases in the empirical line list that we used, or the sparsity of the empirical line list we used. We present our final seven line lists in Tables 2–8 in the Appendix; the tables contain the full lists for those with fewer than 10 lines and only the first five lines for the others. The remainder of those lists are included in machine-readable tables.

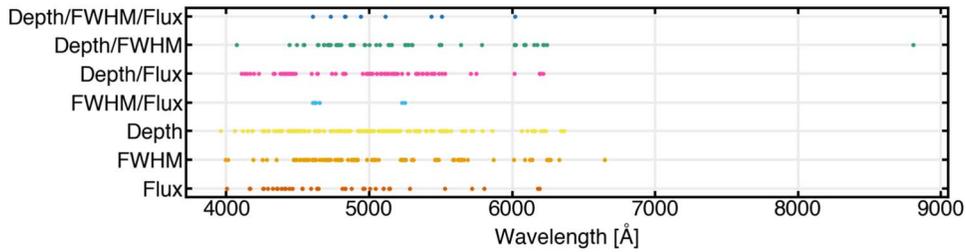

**Figure 5.** Ranges of each of our seven line lists over the full NEID wavelength range (3800–9300 Å). The wavelengths of the lines in each line list are generally uniformly distributed blueward of ~6500 Å, with a notable gap between 6500-9000 Å populated by only a couple of lines.

### 3.3. Interpretation of Physical Behavior of Line Lists

While all seven of our lists contain lines that are correlated with the S-index in some feature and therefore show some sensitivity to stellar activity, each of these individual lists further probes the activity sensitivity of the lines in unique ways and provides additional insight into the mechanisms of how those lines are affected by and trace stellar activity. Here, we offer an initial probe into what those mechanisms may be; we note that we are not aiming to provide a complete theoretical explanation of the physical behaviors of these lines but rather discuss some possible explanations and directions for further research.

In order to investigate on what timescale our correlations may be acting, we generated periodograms of the measured RVs and S-indices for each observation used in our analysis. We used astropy.timeseries.LombScargle (VanderPlas et al. 2012 Astropy Collaboration et al. 2013; VanderPlas & Ivezic 2015; Astropy Collaboration et al. 2018, 2022) to generate these periodograms, shown in Figure 6 along with their time series. Both periodograms exhibit a peak at ~11 days, the rotational period of $\epsilon$ Eri (Howard & Fulton 2016; Mawet et al. 2019). This suggests that the correlations that we see in this paper are tracing stellar activity modulated by the rotational period of the star.

We consider our list of lines that are correlated with the S-index in all three parameters (depth, FWHM, and integrated flux) to be the strongest candidates for activity-sensitive lines. While the other six line lists probe activity sensitivity in unique ways, this line list shows strong correlations in all three of the features we investigated. These lines exhibit variations in depth and FWHM that trace activity and are not only significant enough to independently show correlation with the S-index but also significant enough to concurrently affect integrated flux. Therefore, these lines exhibit shape changes that appear to be the most sensitive to S-index. Notably, when these lines are traced back to their source line list, all are present in multiple ESPRESSO masks for different stellar types. Since we have found these lines to be particularly sensitive to stellar activity, we recommend that these lines be removed from masks used to calculate RVs. Since these lines exhibit concurrent FWHM and integrated flux variation, it is also possible that some of these lines may be "invisibly" blended (i.e., blended lines whose component lines overlap so much they appear to be a single line), and the supposed FWHM variation is actually driven by depth variations in the "invisible" neighboring line. The question of whether any of our lines are blended is investigated in Section 3.4. These FWHM variations could also be due to Zeeman splitting driven by changes in the star's magnetic field strength or asymmetries in the line profile caused by granulation (Gray 2005). The possibility of Zeeman splitting in our lines sensitive to activity in FWHM is further explored in Section 3.4.

Notably, four out of these nine lines were not previously identified by W18 to be activity sensitive; they represent new, independently identified lines that are sensitive to activity in all three of our measured line parameters. Figure 7 shows plots of the correlations between line depth, FWHM, integrated flux, and S-index for these four lines. The other five out of these nine lines are in the list of activity-sensitive lines published by W18, providing a secondary confirmation and strong evidence that these five lines are sensitive to stellar activity. Figure 8 displays plots showing the correlations between each of the three line features and the S-index for these five lines.

The lines that are correlated with the S-index in both depth and FWHM, but not integrated flux, are most likely lines whose depths and FWHMs both vary in tandem with activity in such a way that the integrated flux over the uniform window does not vary significantly. Another explanation is that the window over which we calculated integrated flux is wide enough such that the absolute variations in depth and FWHM are too small to translate into flux variations, although since we use a narrow window that generally captures only the core of each line (see Figure 1), this would be most probable only in lines that are particularly narrow such that most of the line is encompassed by the integrated flux window. Since these lines exhibit FWHM variation, this behavior may also be driven by Zeeman splitting or granulation. The concurrent depth variation may be further evidence that these lines are particularly sensitive to granulation in the photosphere, which we would expect to cause spectral lines to be both shallower *and* broader (Gray 2005). Another feature of this line list is the fact that both the FWHMs and depths of these lines vary with activity, helping to check against the possibility that these lines are tracing some artifact of our continuum normalization scheme in depth variation rather than stellar activity, since the FWHM of the line is less affected by continuum normalization.

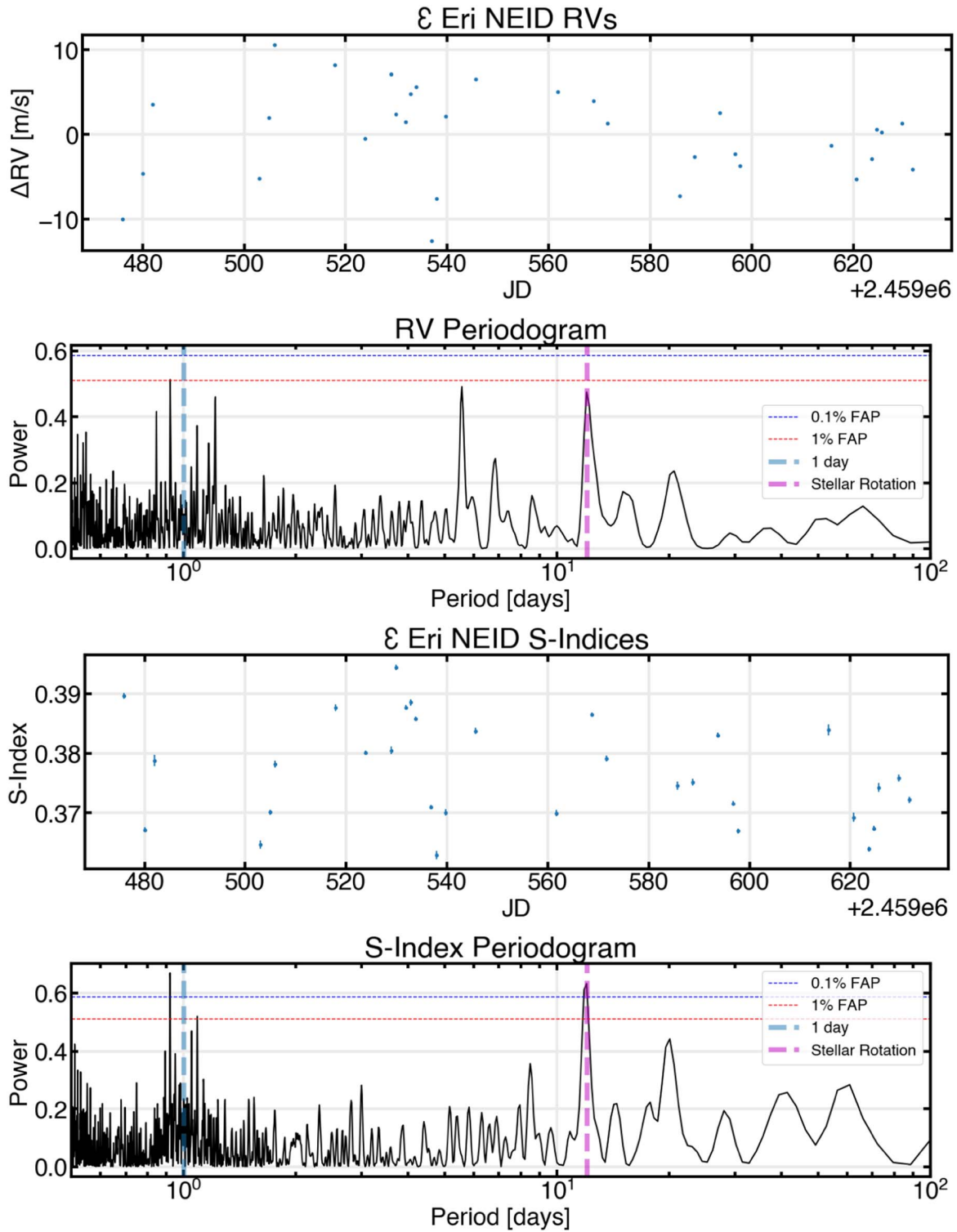

**Figure 6.** Time series of the measured RVs and S-indices and corresponding periodograms for each NEID observation used in our analysis. In the RV time series, we subtract the systematic RV of ε Eri, 16.49 km s$^{-1}$, and plot the Δ RVs. Both periodograms display a peak at ∼11 days, the rotational period of the star (Howard & Fulton 2016; Mawet et al. 2019).

The lines correlated with the S-index in both depth and integrated flux, but not FWHM, are most likely lines in which the variations in integrated flux are driven by variations in depth. The absolute variations of these line depths are large enough to affect the integrated flux as well, which manifests in correlations with the S-index in both parameters. The list of lines correlated with the S-index in both FWHM and integrated flux, but not depth, exhibits similar behavior; since integrated flux is partially tied to FWHM as well, these lines have widths that vary significantly enough with activity to affect the integrated flux. As noted in Section 2.3, because of the narrowness of the window we used to calculate integrated flux, these lines are also further filtered for FWHM variations that are particularly significant enough to affect the core/narrow window. Since these lines do not have depths that also vary similarly with activity, the physical processes that drive variations in the line shape caused by stellar activity in these lines affect mostly width and not depth. As with the other lines that exhibit FWHM variation, some possible physical explanations for this are Zeeman splitting or granulation.

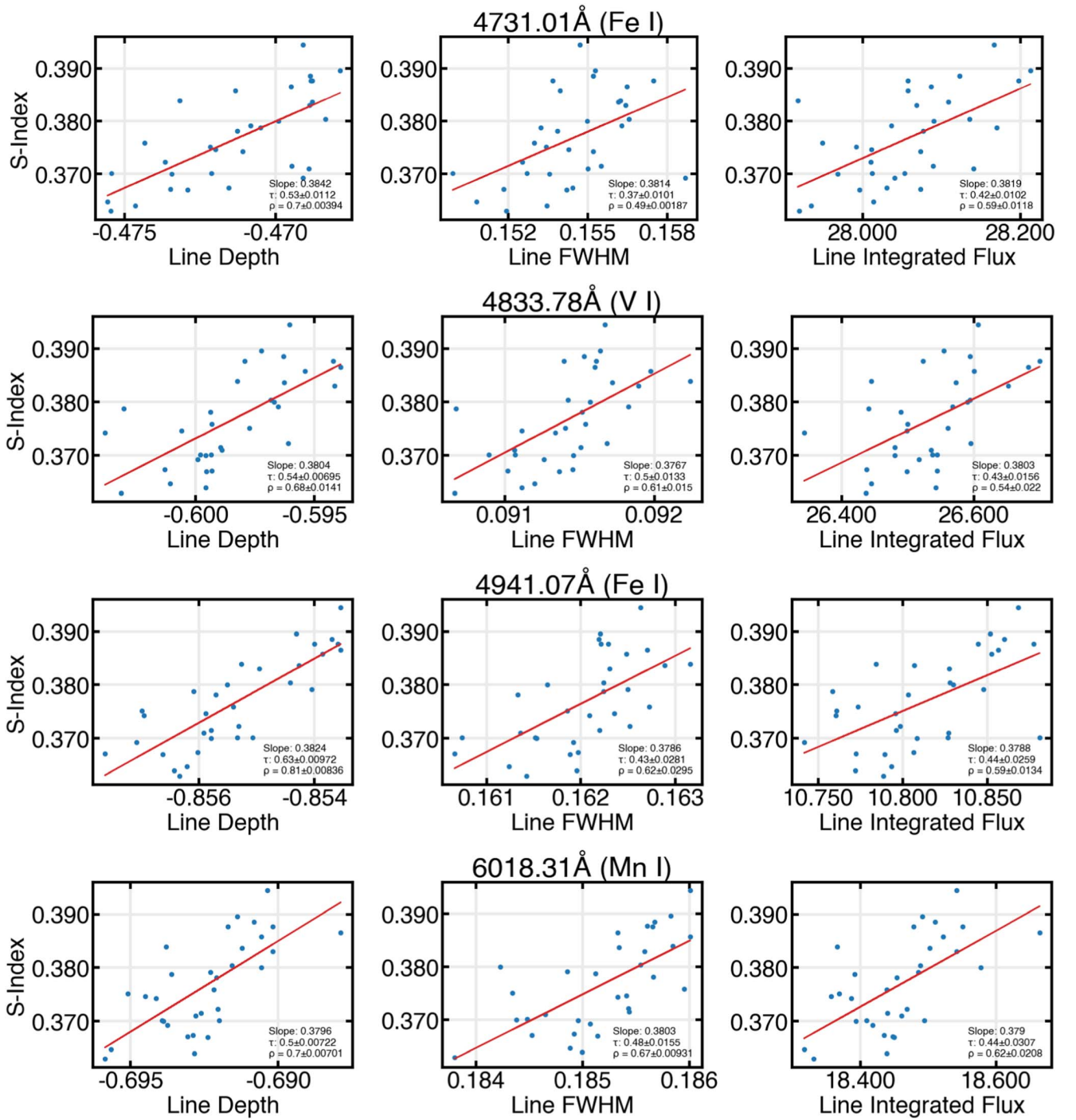

**Figure 7.** Correlations between line depth, FWHM, and integrated flux and S-index for each of the four lines that we found to be strongly correlated with the S-index in all three of our measured line parameters, which do *not* also appear in the line list published by Wise et al. (2018). These lines are newly identified activity-sensitive lines that trace activity in several different line features.

For each of the three lists of lines correlated with the S-index in only one feature, the variability in that feature does not affect the other features significantly enough to be correlated in that feature as well. For example, the lines correlated in depth may only exhibit absolute depth variation that is too small to affect the integrated flux over that line but still vary enough to exhibit correlation with the S-index. Note that these lines are also the most sensitive to the effects of continuum normalization. This is an important distinction from other lines that show a correlation in depth and another feature, where the absolute depth variation is most likely either driving variation in the other parameter (i.e., integrated flux) or at least varying in tandem with it (i.e., FWHM). Similar distinctions apply to the lists of lines correlated in FWHM only and integrated flux only. For the lines correlated in FWHM only, these lines may also exhibit absolute FWHM variation that is too small to affect integrated flux but still varies enough to correlate with activity. As noted earlier, this behavior could be caused by Zeeman splitting or granulation. The lines correlated in integrated flux only, but not FWHM or depth, may exhibit variations in depth and FWHM that are not significant enough to show correlation with the S-index but exert a combined effect on integrated flux that is significant enough to show a correlation.

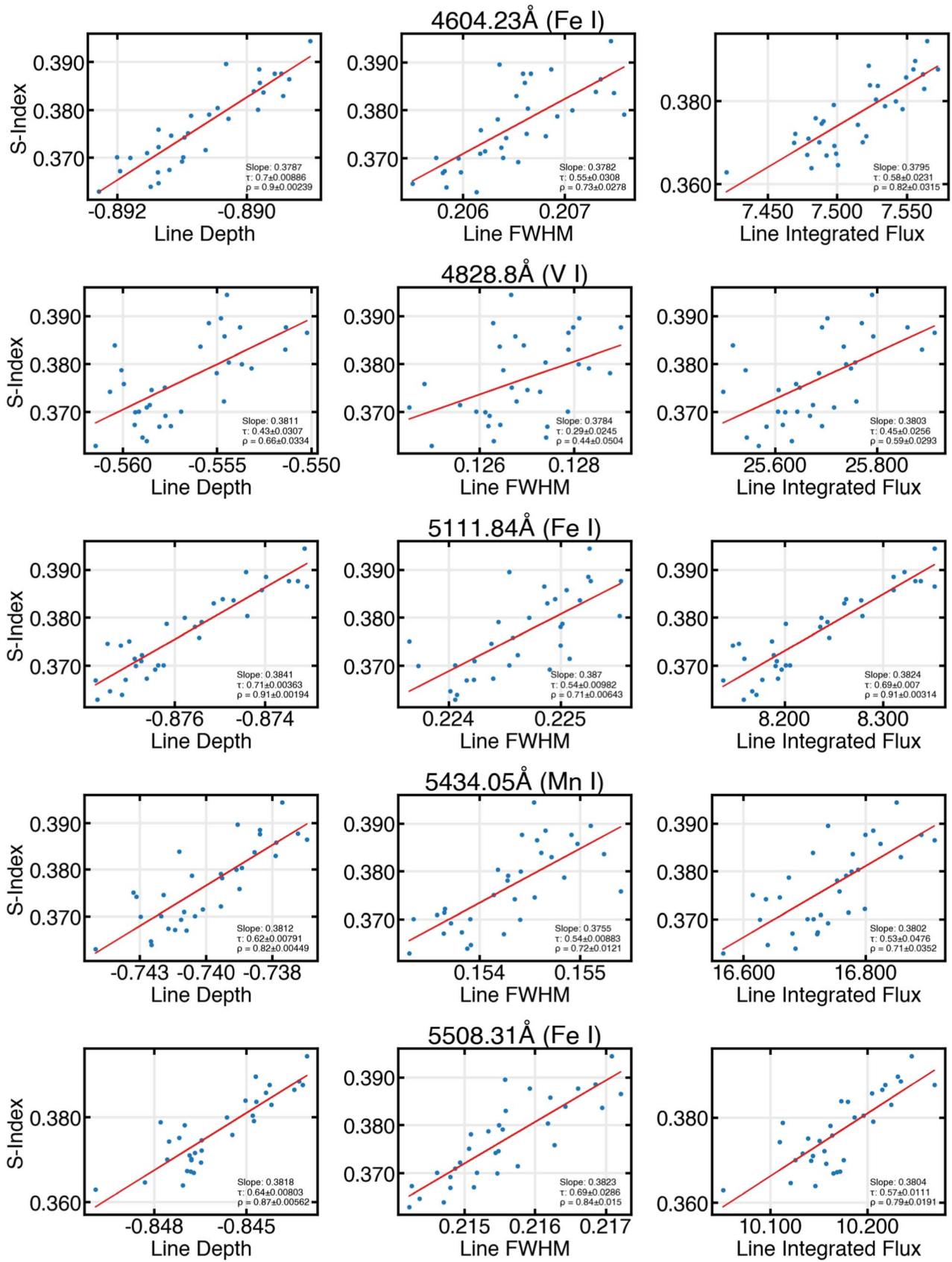

**Figure 8.** Correlations between line depth, FWHM, and integrated flux and S-index for each of the five lines that we found to be strongly correlated with the S-index in all three of our measured line parameters, which also appear in the line list published by Wise et al. (2018). While Wise et al. (2018) found these lines to be correlated with the S-index in line core flux (analogous to line depth), we further confirmed these lines are also sensitive to activity in other parameters (FWHM and integrated flux).

### 3.4. VALD Analysis

As a first step in investigating the physical processes behind the behavior of our lines, we used the line list generated from VALD using the stellar parameters of $\epsilon$ Eri from Drake & Smith (1993) and a line detection threshold of 0.1 to determine the species of our lines, investigate if the activity sensitivity of our lines is correlated with Landé factor, explore the possibility of detecting Zeeman line splitting on our data set, and investigate whether or not any lines in our line lists are blends. The species of each line and the atomic transition with which it is associated are listed in Tables 2–8 in the Appendix.

We explored the relationships between the estimated Landé factor from our VALD query and all $|\tau|$ and $|\rho|$ values for each line across all of our activity-sensitive lines. These relationships measure the dependence of the strength of the line's sensitivity to activity on the line's Landé factor, and we calculate $|\tau|$ and $|\rho|$ values to quantify each relationship. Similar to Wise et al. (2022), we found no correlation coefficient >0.2 between Landé factor and any $|\tau|$ or $|\rho|$. This suggests that the strength of a line's sensitivity to activity is not solely dependent on its Landé factor.

We also used VALD to explore the possibility of empirically detecting Zeeman line splitting in our data set, which would enable a physically motivated identification of active lines. To do this, we scrutinized the highest FWHM lines in our data cube, since that is how evidence of line splitting would be captured in automated routines fitting single Gaussians. These lines are typically rejected under suspicion of being blends. However, correlating (1) all the lines and (2) only the 50 lines with the widest FWHMs from our line lists with VALD atomic properties yielded no significant ($|\rho| > 0.5$, per our threshold) correlation with any properties. Our conclusion, consistent with extant literature, is that the Zeeman signature is difficult to directly measure on individual lines, at these resolutions, on the unpolarized disk-integrated profile of a solar-type star (Semel et al. 2009).

We then investigated our lines for blends. We considered a line to be "invisibly blended" if we found one or more VALD lines within the same $\pm 3$ km s$^{-1}$ window of the line center that we used to calculate integrated flux, a method similar to the one employed by Cretignier et al. (2020) which used a depth contamination threshold to detect blended spectral lines. In order to be self-consistent, we search for contaminants specifically within the same window across which we measured the integrated flux since we expect correlated effects to be contained in that region. The results of this analysis are listed in the tables of our line lists in the Appendix. Notably, we found that none of the lines correlated with the S-index in FWHM and integrated flux only are "invisible" blends, suggesting that the FWHM variation in these lines is driven by true width variation, but several of our other lines correlated with the S-index in other parameters did exhibit one or more VALD lines within $\pm 3$ km s$^{-1}$ of their line centers, including two out of our nine lines correlated with the S-index in all three parameters. Figure 9 plots these two lines along with the VALD wavelengths and depths of the lines with which they are blended. This provides further evidence that the aforementioned lines should not be used for RV measurement.

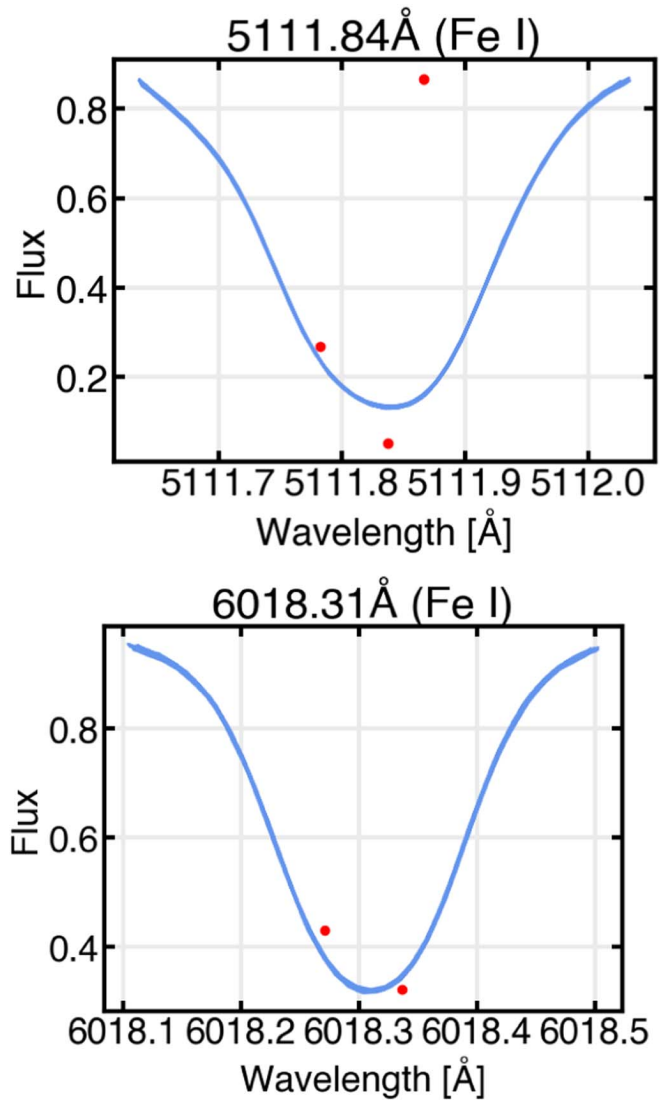

**Figure 9.** Two lines that we found to be correlated with the S-index in all three of our line parameters and that have known blends within $\pm 3$ km s$^{-1}$ of the line center according to the Vienna Atomic Line Database (VALD). The positions and depths of the lines extracted from VALD are shown in red.

### 3.5. Removing Activity-sensitive Lines from ESPRESSO Masks

As an initial test of how the removal of the lines we identified to be the most sensitive to activity affects measured RVs, we computed the RVs of our NEID $\epsilon$ Eri observations using three separate masks: (1) the combined ESPRESSO mask that we used in our data cube, (2) just the G2 ESPRESSO mask, and (3) just the K2 ESPRESSO mask. For each mask, we also created a corresponding mask with our most activity-sensitive lines removed. Comparing the RVs measured using the full masks versus the masks with the lines removed resulted in improvements of 2–3 cm s$^{-1}$. While the improvement is small in this data set, this result further cautions against the use of these lines to measure RVs.

## 4. Conclusion

In this paper, we present a method for analyzing stellar spectra and correlating line features with stellar activity indicators, apply this method to NEID $\epsilon$ Eri data, and filter the results to find well-defined lines correlated with the S-index in one or more line properties. We further confirmed the list of activity-sensitive lines published by W18, particularly five lines that exhibited strong sensitivity to activity in all three of the line features we investigated. We present a list of lines correlated with the S-index in three different line features (depth, FWHM, and integrated flux) as well as six additional lists of lines correlated with the S-index in one or two line features, each of which probes the sensitivity of those particular lines to stellar activity in unique ways and offers further avenues of exploration into the specific physical processes and types of stellar activity that each line traces. Our strongest list of activity-correlated lines, correlated with the S-index in depth, FWHM, and integrated flux, is a useful reference to inform future stellar activity studies and include four newly identified lines sensitive to activity in all three of our measured line parameters. As all nine of our strongest list of activity-correlated lines were sourced from the combined ESPRESSO mask list, and each appears in multiple ESPRESSO masks, we recommend that these lines be removed from these masks and not used to calculate RVs.

While we did not exhaustively explore the physical processes behind each line list that might inform why certain lines are correlated in some features and not others in this paper, we present some possible explanations for the behavior we observed and present our line lists as starting points for future investigations into the processes in the star that affect these spectral lines in specific ways. Some possible directions for future research include:

1. Further investigating the physical processes that cause variations in particular line features and not others, and therefore providing further insight into the types of stellar activity that different lines trace.
2. Further probing the wavelength region between 6500 and 9000 Å specifically to investigate the effect(s) of telluric bands/other physical factors on lines in this region and help explain the difficulty of identifying activity-sensitive lines in this region.
3. Applying our analysis to and/or using these line lists as references of activity-sensitive lines for further studies on other stars, different features, and/or different activity indicators.
4. Probing the periodicities of the features of the lines presented here in larger data sets of $\epsilon$ Eri and/or other stars to further confirm the sensitivity of these lines to rotationally modulated activity.

These suggestions for potential future research would allow greater precision in using the variation in these lines and features to identify and separate different types of stellar activity from radial velocity measurements. Understanding the relationship between the spectral features and the physical processes driving stellar activity in stars, and how they affect radial velocity measurements, is crucial for increasing the precision and sensitivity of EPRV measurements to small, Earth-sized exoplanets. Particularly, with the potential of machine learning to open new avenues in stellar activity mitigation and improve the precision of EPRV measurements, building a foundation of knowledge of the physical relationships between spectra and stellar activity is crucial to informing the design and interpretation of the results of these models. Our line lists presented in this paper offer a first probe into demystifying these relationships and a useful starting point for future studies that will help pave the way to improving the RV method for detecting Earth-sized and potentially habitable exoplanets.

## Acknowledgments

Data presented were obtained by the NEID spectrograph built by Penn State University and operated at the WIYN Observatory by NOIRLab, under the NN-EXPLORE partnership of the National Aeronautics and Space Administration (NASA) and the National Science Foundation (NSF). This work has also made use of the VALD database, operated at Uppsala University, the Institute of Astronomy RAS in Moscow, and the University of Vienna. This research was, in part, carried out at the Jet Propulsion Laboratory (JPL), California Institute of Technology, under a contract with NASA (80NM0018D0004). S.J. and A.R. acknowledge support from the Space Telescope Science Institute (STScI) Director's Discretionary Fund, the NASA/JPL NEID GO program, and the JPL R&TD Fund. C.F.B. acknowledges support for NEID pipeline development and analysis under JPL subcontract 1644767. S.J. thanks Stephanie LaMassa for useful advice and support during this project. We also thank Andrew Howard for useful conversations on this project.

*Facility:* WIYN (NEID).

*Software*: astropy (Astropy Collaboration et al. 2013, 2018, 2022), scipy (Virtanen et al. 2020), xarray (Hoyer & Hamman 2017).

# Appendix
# Tables of Line Lists

The following tables (Tables 2–8) contain the lists of lines correlated with the S-index in one or more parameters along with their correlations, species and atomic transition with which it is associated, and whether or not the line is blended. The tables contain the full list for those with fewer than ten lines and only the first five lines for the others, the remainder of which are included in machine-readable tables.

**Table 2**
Depth/FWHM/Integrated Flux-correlated Lines

| Line (Å) | Depth $|\rho|$ | FWHM $|\rho|$ | Flux $|\rho|$ | Depth $|\tau|$ | FWHM $|\tau|$ | Flux $|\tau|$ | Species | Blend? |
|---|---|---|---|---|---|---|---|---|
| 4604.23 | 0.90 ± 0.002 | 0.73 ± 0.03 | 0.82 ± 0.03 | 0.70 ± 0.009 | 0.55 ± 0.003 | 0.58 ± 0.02 | Fe I | N |
| 4731.01 | 0.70 ± 0.004 | 0.49 ± 0.002 | 0.59 ± 0.01 | 0.53 ± 0.01 | 0.37 ± 0.01 | 0.42 ± 0.01 | Fe I | N |
| 4828.80 | 0.66 ± 0.03 | 0.44 ± 0.05 | 0.59 ± 0.03 | 0.43 ± 0.03 | 0.29 ± 0.03 | 0.45 ± 0.03 | V I | N |
| 4833.78 | 0.68 ± 0.01 | 0.61 ± 0.02 | 0.54 ± 0.02 | 0.54 ± 0.007 | 0.50 ± 0.01 | 0.43 ± 0.02 | V I | N |
| 4941.07 | 0.81 ± 0.008 | 0.62 ± 0.03 | 0.59 ± 0.01 | 0.63 ± 0.01 | 0.43 ± 0.03 | 0.44 ± 0.03 | Fe I | N |
| 5111.84 | 0.91 ± 0.002 | 0.71 ± 0.006 | 0.91 ± 0.003 | 0.71 ± 0.004 | 0.54 ± 0.01 | 0.69 ± 0.007 | Fe I | Y |
| 5434.05 | 0.82 ± 0.005 | 0.72 ± 0.01 | 0.71 ± 0.04 | 0.62 ± 0.008 | 0.54 ± 0.009 | 0.53 ± 0.05 | Mn I | N |
| 5508.31 | 0.87 ± 0.006 | 0.84 ± 0.02 | 0.79 ± 0.02 | 0.64 ± 0.008 | 0.69 ± 0.03 | 0.57 ± 0.01 | Fe I | N |
| 6018.31 | 0.70 ± 0.007 | 0.67 ± 0.009 | 0.62 ± 0.02 | 0.50 ± 0.007 | 0.48 ± 0.02 | 0.44 ± 0.03 | Mn I | Y |

**Note.** Wavelengths reported are wavelengths in vacuum and in the stellar rest frame.

(This table is available in its entirety in machine-readable form.)

**Table 3**
Depth/FWHM-correlated Lines

| Line (Å) | Depth $|\rho|$ | FWHM $|\rho|$ | Flux $|\rho|$ | Depth $|\tau|$ | FWHM $|\tau|$ | Flux $|\tau|$ | Species | Blend? |
|---|---|---|---|---|---|---|---|---|
| 4071.43 | 0.64 ± 0.002 | 0.57 ± 0.007 | 0.26 ± 0.03 | 0.48 ± 0.006 | 0.41 ± 0.009 | 0.25 ± 0.03 | Mn I | N |
| 4439.59 | 0.70 ± 0.01 | 0.51 ± 0.03 | 0.34 ± 0.05 | 0.49 ± 0.007 | 0.42 ± 0.009 | 0.18 ± 0.05 | Fe I | N |
| 4491.00 | 0.70 ± 0.01 | 0.77 ± 0.002 | 0.47 ± 0.02 | 0.56 ± 0.009 | 0.59 ± 0.006 | 0.34 ± 0.02 | Fe I | N |
| 4541.98 | 0.76 ± 0.14 | 0.74 ± 0.07 | 0.11 ± 0.08 | 0.57 ± 0.11 | 0.53 ± 0.07 | 0.08 ± 0.06 | Cr I | Y |
| 4542.34 | 0.53 ± 0.04 | 0.72 ± 0.06 | 0.01 ± 0.05 | 0.43 ± 0.04 | 0.54 ± 0.04 | 0.03 ± 0.04 | Cr I | N |

**Note.** The full version of this table is available in machine-readable format. A selection of lines is previewed here. Wavelengths reported are wavelengths in vacuum and in the stellar rest frame.

(This table is available in its entirety in machine-readable form.)

**Table 4**
Depth/Integrated Flux-correlated Lines

| Line (Å) | Depth $|\rho|$ | FWHM $|\rho|$ | Flux $|\rho|$ | Depth $|\tau|$ | FWHM $|\tau|$ | Flux $|\tau|$ | Species | Blend? |
|---|---|---|---|---|---|---|---|---|
| 4104.10 | 0.65 ± 0.01 | 0.00 ± 0.03 | 0.61 ± 0.03 | 0.47 ± 0.006 | 0.00 ± 0.02 | 0.46 ± 0.03 | Si I | N |
| 4129.23 | 0.63 ± 0.08 | 0.20 ± 0.09 | 0.57 ± 0.03 | 0.50 ± 0.07 | 0.15 ± 0.04 | 0.42 ± 0.03 | V I | N |
| 4145.04 | 0.73 ± 0.03 | 0.11 ± 0.006 | 0.74 ± 0.08 | 0.53 ± 0.04 | 0.06 ± 0.008 | 0.55 ± 0.07 | Fe I | N |
| 4167.49 | 0.65 ± 0.02 | 0.15 ± 0.07 | 0.57 ± 0.08 | 0.50 ± 0.02 | 0.09 ± 0.05 | 0.46 ± 0.06 | Ti I | N |
| 4191.89 | 0.70 ± 0.02 | 0.03 ± 0.03 | 0.71 ± 0.07 | 0.51 ± 0.009 | 0.02 ± 0.02 | 0.52 ± 0.05 | Co I | N |

**Note.** The full version of this table is available in machine-readable format. A selection of lines is previewed here. Wavelengths reported are wavelengths in vacuum and in the stellar rest frame.

(This table is available in its entirety in machine-readable form.)

**Table 5**
FWHM/Integrated Flux-correlated Lines

| Line (Å) | Depth $|\rho|$ | FWHM $|\rho|$ | Flux $|\rho|$ | Depth $|\tau|$ | FWHM $|\tau|$ | Flux $|\tau|$ | Species | Blend? |
|---|---|---|---|---|---|---|---|---|
| 4603.29 | 0.04 ± 0.005 | 0.65 ± 0.02 | 0.49 ± 0.02 | 0.03 ± 0.008 | 0.48 ± 0.008 | 0.32 ± 0.02 | Fe I | N |
| 4617.42 | 0.45 ± 0.02 | 0.85 ± 0.008 | 0.61 ± 0.08 | 0.33 ± 0.02 | 0.68 ± 0.007 | 0.44 ± 0.07 | Cr I | N |
| 4627.47 | 0.27 ± 0.008 | 0.71 ± 0.02 | 0.46 ± 0.02 | 0.16 ± 0.01 | 0.54 ± 0.02 | 0.33 ± 0.03 | Cr I | N |
| 4652.59 | 0.00 ± 0.09 | 0.64 ± 0.03 | 0.61 ± 0.07 | 0.02 ± 0.04 | 0.47 ± 0.02 | 0.42 ± 0.03 | Cr I | N |
| 5226.98 | 0.57 ± 0.02 | 0.92 ± 0.003 | 0.68 ± 0.04 | 0.40 ± 0.01 | 0.75 ± 0.008 | 0.50 ± 0.04 | Fe I | N |
| 5248.51 | 0.56 ± 0.04 | 0.91 ± 0.005 | 0.54 ± 0.02 | 0.39 ± 0.01 | 0.74 ± 0.02 | 0.39 ± 0.01 | Fe I | N |

**Note.** Wavelengths reported are wavelengths in vacuum and in the stellar rest frame.

(This table is available in its entirety in machine-readable form.)

Table 6
Depth-correlated Lines

| Line (Å) | Depth $|\rho|$ | FWHM $|\rho|$ | Flux $|\rho|$ | Depth $|\tau|$ | FWHM $|\tau|$ | Flux $|\tau|$ | Species | Blend? |
|---|---|---|---|---|---|---|---|---|
| 3959.33 | 0.52 ± 0.02 | 0.41 ± 0.03 | 0.33 ± 0.07 | 0.43 ± 0.02 | 0.31 ± 0.03 | 0.25 ± 0.02 | Ti I | Y |
| 4056.69 | 0.60 ± 0.04 | 0.51 ± 0.02 | 0.34 ± 0.2 | 0.47 ± 0.03 | 0.31 ± 0.02 | 0.22 ± 0.01 | Mn I | Y |
| 4059.35 | 0.65 ± 0.02 | 0.30 ± 0.01 | 0.51 ± 0.11 | 0.46 ± 0.03 | 0.22 ± 0.02 | 0.33 ± 0.07 | Fe I | Y |
| 4113.87 | 0.19 ± 0.003 | 0.21 ± 0.002 | 0.15 ± 0.02 | 0.26 ± 0.06 | 0.20 ± 0.08 | 0.10 ± 0.04 | Ti I | N |
| 4148.16 | 0.62 ± 0.08 | 0.41 ± 0.07 | 0.25 ± 0.07 | 0.48 ± 0.08 | 0.27 ± 0.06 | 0.18 ± 0.05 | Fe I | N |

**Note.** The full version of this table is available in machine-readable format. A selection of lines is previewed here. Wavelengths reported are wavelengths in vacuum and in the stellar rest frame.

(This table is available in its entirety in machine-readable form.)

Table 7
FWHM-correlated Lines

| Line (Å) | Depth $|\rho|$ | FWHM $|\rho|$ | Flux $|\rho|$ | Depth $|\tau|$ | FWHM $|\tau|$ | Flux $|\tau|$ | Species | Blend? |
|---|---|---|---|---|---|---|---|---|
| 3992.56 | 0.47 ± 0.006 | 0.50 ± 0.02 | 0.08 ± 0.12 | 0.41 ± 0.02 | 0.43 ± 0.006 | 0.02 ± 0.11 | Co I | N |
| 4012.54 | 0.04 ± 0.1 | 0.58 ± 0.01 | 0.47 ± 0.07 | 0.09 ± 0.06 | 0.40 ± 0.01 | 0.31 ± 0.04 | Fe I | Y |
| 4185.48 | 0.34 ± 0.03 | 0.53 ± 0.01 | 0.14 ± 0.08 | 0.22 ± 0.008 | 0.41 ± 0.03 | 0.17 ± 0.06 | Ti II | N |
| 4251.32 | 0.02 ± 0.02 | 0.54 ± 0.006 | 0.38 ± 0.02 | 0.04 ± 0.02 | 0.41 ± 0.009 | 0.27 ± 0.02 | Fe I | N |
| 4282.58 | 0.40 ± 0.1 | 0.63 ± 0.04 | 0.25 ± 0.09 | 0.28 ± 0.09 | 0.44 ± 0.03 | 0.19 ± 0.06 | Ti I | N |

**Note.** The full version of this table is available in machine-readable format. A selection of lines is previewed here. Wavelengths reported are wavelengths in vacuum and in the stellar rest frame.

(This table is available in its entirety in machine-readable form.)

Table 8
Integrated Flux-correlated Lines

| Line (Å) | Depth $|\rho|$ | FWHM $|\rho|$ | Flux $|\rho|$ | Depth $|\tau|$ | FWHM $|\tau|$ | Flux $|\tau|$ | Species | Blend? |
|---|---|---|---|---|---|---|---|---|
| 4001.59 | 0.43 ± 0.08 | 0.34 ± 0.06 | 0.56 ± 0.05 | 0.40 ± 0.08 | 0.25 ± 0.06 | 0.40 ± 0.04 | Fe I | Y |
| 4161.95 | 0.42 ± 0.07 | 0.22 ± 0.04 | 0.57 ± 0.09 | 0.31 ± 0.05 | 0.20 ± 0.04 | 0.42 ± 0.07 | Fe I | N |
| 4164.83 | 0.44 ± 0.05 | 0.12 ± 0.004 | 0.52 ± 0.07 | 0.28 ± 0.03 | 0.17 ± 0.008 | 0.34 ± 0.06 | Fe I | Y |
| 4255.54 | 0.23 ± 0.02 | 0.43 ± 0.02 | 0.61 ± 0.03 | 0.18 ± 0.03 | 0.33 ± 0.008 | 0.47 ± 0.04 | Cr I | N |
| 4261.68 | 0.58 ± 0.02 | 0.08 ± 0.009 | 0.65 ± 0.04 | 0.40 ± 0.02 | 0.03 ± 0.008 | 0.47 ± 0.02 | Fe I | N |

**Note.** The full version of this table is available in machine-readable format. A selection of lines is previewed here. Wavelengths reported are wavelengths in vacuum and in the stellar rest frame.

(This table is available in its entirety in machine-readable form.)

## ORCID iDs

Sarah Jiang https://orcid.org/0000-0002-4406-2727
Arpita Roy https://orcid.org/0000-0001-8127-5775
Samuel Halverson https://orcid.org/0000-0003-1312-9391
Chad F. Bender https://orcid.org/0000-0003-4384-7220
O. Justin Otor https://orcid.org/0000-0002-4679-5692
Suvrath Mahadevan https://orcid.org/0000-0001-9596-7983
Guðmundur Stefánsson https://orcid.org/0000-0001-7409-5688
Ryan C. Terrien https://orcid.org/0000-0002-4788-8858
Christian Schwab https://orcid.org/0000-0002-4046-987X

## References

Altrock, R. C., November, L. J., Simon, G. W., Milkey, R. W., & Worden, S. P. 1975, SoPh, 43, 33
Astropy Collaboration, Price-Whelan, A. M., Lim, P. L., et al. 2022, ApJ, 935, 167
Astropy Collaboration, Price-Whelan, A. M., Sipőcz, B. M., et al. 2018, AJ, 156, 123
Astropy Collaboration, Robitaille, T. P., & Tollerud, E. J. 2013, A&A, 558, A33
Blackman, R. T., Fischer, D. A., Jurgenson, C. A., et al. 2020, AJ, 159, 238
Cretignier, M., Dumusque, X., Allart, R., Pepe, F., & Lovis, C. 2020, A&A, 633, A76
Cretignier, M., Dumusque, X., & Pepe, F. 2022, A&A, 659, A68
de Beurs, Z. L., Vanderburg, A., Shallue, C. J., et al. 2022, AJ, 164, 49
Drake, J. J., & Smith, G. 1993, ApJ, 412, 797
Ervin, T., Halverson, S., Burrows, A., et al. 2022, AJ, 163, 272
Gibson, S. R., Howard, A. W., Roy, A., et al. 2018, Proc. SPIE, 10702, 107025X
Gray, D. F. 2005, The Observation and Analysis of Stellar Photospheres (Cambridge: Cambridge Univ. Press)
Haywood, R. D., Milbourne, T. W., Saar, S. H., et al. 2022, ApJ, 935, 6
Howard, A. W., & Fulton, B. J. 2016, PASP, 128, 114401
Hoyer, S., & Hamman, J. 2017, JOSS, 5, 10
Isaacson, H., & Fischer, D. 2010, ApJ, 725, 875
Jeffers, S. V., Barnes, J. R., Schöfer, P., et al. 2022, A&A, 663, A27
Lisogorskyi, M., Jones, H. R. A., Feng, F., Butler, R. P., & Vogt, S. 2020, MNRAS, 500, 548
Mawet, D., Hirsch, L., Lee, E. J., et al. 2019, AJ, 157, 33
National Academies of Sciences, Engineering, and Medicine 2018, Exoplanet Science Strategy (Washington, DC: The National Academies Press)
Pakhomov, Y. V., Ryabchikova, T. A., & Piskunov, N. E. 2019, ARep, 63, 1010
Pepe, F., Cristiani, S., Rebolo, R., et al. 2021, A&A, 645, A96

Pepe, F., Mayor, M., Delabre, B., et al. 2000, Proc. SPIE, 4008, 582
Robertson, P., Mahadevan, S., Endl, M., & Roy, A. 2014, Sci, 345, 440
Sairam, L., & Triaud, A. H. M. J. 2022, MNRAS, 514, 2259
Schwab, C., Rakich, A., Gong, Q., et al. 2016, Proc. SPIE, 9908, 99087H
Seifahrt, A., Bean, J. L., Stürmer, J., et al. 2020, Proc. SPIE, 11447, 114471F
Semel, M., Ramírez Vélez, J. C., Martínez González, M. J., et al. 2009, A&A, 504, 1003
Siegel, J. C., Rubenzahl, R. A., Halverson, S., & Howard, A. W. 2022, AJ, 163, 260
Simola, U., Bonfanti, A., Dumusque, X., et al. 2022, A&A, 664, A127
VanderPlas, J., Connolly, A. J., Ivezic, Z., & Gray, A. 2012, in 2012 Conference on Intelligent Data Understanding (New York: IEEE), 47
VanderPlas, J. T., & Ivezic, Z. 2015, ApJ, 812, 18
Vaughan, A. H., Preston, G. W., & Wilson, O. C. 1978, PASP, 90, 267
Virtanen, P., Gommers, R., Oliphant, T. E., et al. 2020, NatMe, 17, 261
Wilson, O. C. 1968, ApJ, 153, 221
Wise, A., Plavchan, P., Dumusque, X., Cegla, H., & Wright, D. 2022, ApJ, 930, 121
Wise, A. W., Dodson-Robinson, S. E., Bevenour, K., & Provini, A. 2018, AJ, 156, 180
Zhao, J., Ford, E. B., & Tinney, C. G. 2022, ApJ, 935, 75